# The Emotion coding and Propagation based on improved Genetic algorithm

Hongyuan Diao, Fuzhong Nian*, Xuelong Yu, Xirui Liu and Xinhao Liu

.

*Abstract*—Computational communication research on information has been prevalent in recent years, as people are progressively inquisitive in social behavior and public opinion. Nevertheless, it is of great significance to analyze the direction of predominant sentiment from the sentiment communication perspective. In this paper, the information emotion propagation model is established by introducing revamp genetic algorithms into information emotion. In the process of information dissemination, both the information emotions and the network emotions are dynamic. For this model, the information emotions and the network nodes emotions are quantified as binary codes. The convergence effects, crossover and mutation algorithms are introduced. These factors all act on the transmission process via dynamic propagation rate, and the improved genetic algorithm also acts on the emotion transmission. In particular, the latter two algorithms are different from the existing biological domain. Based on the existing research results in other manuscripts, we perform simulation described above on the hybrid network. The simulation results demonstrate that the trend approximate to the actual data. As a result, our work can prove that our proposed model is essentially consistent with the actual emotion transmission phenomenon.

## I. Introduction

INFORMATION propagation is important research field in social networks [1-15]. Researching the spread phenomenon help us better understand the social. The early works are Barabási and Albert proposed the free scale [16], and Watts and Strogatz proposed the small world network model [17]. Then, the epidemiological models proposed, for example, the susceptible-infected model (SI model)[18], the susceptible-infected-recovered model (SIR model) [19, 20] and susceptible-infected-susceptible model (SIS model) [21, 22]. The classic D-K model [23] adopts the stochastic process method to analyze rumors and divides the audience into three categories in accordance with the rumors pervasion: the ignorants, spreaders and stiflers. In addition, the three categories corresponding to the three populations in the SIR model [19, 24].

In recent years, domestic and foreign research in information emotion has distributed in many fields, such as journalism, sociology and computational communication. Research in multiple fields provides different perspectives on this issue. Such as, a new approach of genetic programming for music emotion classification is suggested [25]. Chen *et al*. studied about "storage and management strategy for heterogeneous data stream based on mutation information" [26]. The study is sentiment analysis in Twitter text [27]. The research is about "estimation of information via variation" [28]. An information theory method to study the effect of mutation on genetic algorithms is used [29]. The interaction between the volume of news and the intensity of emotions on emerging infectious diseases was studied by investigating Weibo data. Mainly rely on impulse response function for research [30]. Wang *et al*. proposed emotion-based independent cascade model of emotion communication [31]. Introduced a graphically coupled Hidden Markov Emotion Model to simulate and infer the spread of negative emotions of mobile phone users in social networks [32]. Zhang *et al*. studied that based on the Multiagent method, a nonlinear emergence model of the fragility of the social consensus system is established, and the evolution mechanism of the social consensus system for the spread of negative

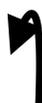

This research is supported by the National Natural Science Foundation of China (No. 61863025), Pro-gram for International S & T Cooperation Projects of Gansu Province (No. 144WCGA166), Program for Longyuan Young Innovation Talents and the Doctor-al Foundation of LUT.

Hongyuan Diao is with the School of Computer and Communication, Lan-zhou University of Technology, LanZhou. E-mail: hydiao.yui@foxmail.com.

*Corresponding author: Fuzhong Nian is with the School of Computer and Communication, Lan-zhou University of Technology, LanZhou, 730050 China. E-mail: gdnfz@lut.cn.

Xuelong Yu, Xirui Liu and Xinhao Liu are with the School of Computer and Communication, Lanzhou University of Technology, LanZhou

emotions [33]. Some of the above researchers apply genetic algorithms [34-37] as a method of cogitation to research the transmutation in emotions, and others apply novel models to research the propagation of emotions. Nevertheless, those have not considered that the interaction between the information emotion and the vertexes emotion in the network. Among them, the carrier is not only include the information text, in essence the emotion in information is the main in the propagation. For example, few researchers employ the method to the emotion propagation. Our principal ideas are introduced emotion-related research based on the genetic algorithm, the influence of emotion propagation on the nodes and the nodes emotion on the propagation process are quantified. We synthesize the above viewpoints and establish the model of information and emotion propagation. So as to verify the rationality of the model better, it is selected that the polls collected by the Los Angeles Times during the "2016 US General Election" as the reference data. In accordance with the voting mechanism of public opinion surveys, our emotional propagation model is eventually manifested in the form of voting. The main contributions of this work are as follows. (1) we quantitate the emotion by the binary system code; (2) we improved the classical the genetic algorithm to drive the binary system emotion code iteration; (3) we introduced a new emotion spread model (SI'I model) according to the improved genetic algorithm; (4) simulation of voting in accordance with the "2016 US General Election".

The organization follows as: (1) in Section 2, the genetic algorithm of emotional coding is introduced in emotion propagation dynamic model. (2) In Section 3: The numerical simulation is finished. The data-driven simulation of single and multi-information is carried out based on the emotion propagation dynamic model.

## II. Model Definition

### A. Review Stage Emotion Coding

The transformations of the information view(opinion) and node emotion always accompanied by that the information spread in network. In particular, the information view(opinion) of one event have different fragments in special time, so that the information fragment take the network different affect. The coupling phenomenon exist between different information fragments which born in the event development, e.g. the similar information fragments view(opinion) facilitate the spread each other and vice versa. In other words, if the node's emotion affected by the information fragment's view(opinion) after had received it, then the other similar view(opinion) of information fragments will influence the node more easily.

The mathematical model is explained from micro perspective, the node emotion will be influenced by how much gap the information fragment's view(opinion) and node emotion have, when the node receives an information fragment. In other words, people will have its own opinion when he knows a massage. After many information fragments spread in network one by one, the network's emotion also has changed many times. From micro perspective, the description about the model is that the network group's emotion gradually become different with beginning time, cause each node's emotion is influenced by network group's emotion and neighbor nodes' emotion.

Under the network topology, the users, the connection of users and the weighted connections are abstractive defined as set $V$ and set $E$ and $\omega \in \Omega$ , $\omega \sim N(\mu, \sigma^2)$ , respectively. Additional, the network topology is signified as $G = (V, E, \Omega)$.

Equation (1) is defined as the total rounds $T$ of spread,

$$T = [0, 1, \cdots, t] \tag{1}$$

where $t$ indicates the round of spread.

In the spread, the model sets that information fragments with different emotion totally have $n$ pieces. Multiple information fragments form a complete event. Under network, information fragment $info_n$ totally spread $t_{info_n}$ rounds in the period $T_{info_n}$. The specific definition is as follows:

$$T = T_{info_1} + T_{info_2} + \cdots + T_{info_n}. \tag{2}$$

Each period is accumulated. We get (3)

$$
\begin{aligned}
T = &[0_{info_1}, 1_{info_1}, 2_{info_1}, \cdots, t_{info_1}] \\
&+ [0_{info_2}, 1_{info_2}, 2_{info_2}, \cdots, t_{info_2}] + \cdots \\
&+ [0_{info_n}, 1_{info_n}, 2_{info_n}, \cdots, t_{info_n}].
\end{aligned} \tag{3}
$$

From (3), we get the complete constituting expression of time sequence, as follows:

$$
T = \begin{bmatrix}
0_{info_1}, 1_{info_1}, 2_{info_1}, \cdots, t_{info_1}, \\
0_{info_2}, 1_{info_2}, 2_{info_2}, \cdots, t_{info_2}, \\
\cdots, \\
0_{info_n}, 1_{info_n}, 2_{info_n}, \cdots, t_{info_n}
\end{bmatrix}. \tag{4}
$$

The expressions of segmented timelines constitute the entire timeline, as shown in (1), which is used to count that the effects executing on the whole process, including the mutation mechanism, the convergence effect and the mapping of the entire process statistics described later. The segmented timeline is defined as in (4), which is used to segment the executing time of which effects in information propagation, and the (4) is defined for the Emotional Similarity Evaluation Function (ESWF) that will be introduced later.

When each node $i$ spreads in time $t$, the node emotion code $E\_code_i(t)$ with $m$-bit $q_m$ (value 0 or 1) is defined as (6), m is an even number, and its Emotional Tendency Value (ETV) is defined as (7), $E_i(t)$ represents the total number of 1 in $E_{code_i}(t)$ code.

$$q_m = \begin{cases} 0 \\ 1 \end{cases} (m > 0) \tag{5}$$

$$E_{code_i}(t) = [q_1, q_2, \cdots, q_m] \tag{6}$$

such as $E_{code_i}(t) = [\underbrace{0,1,1,0,0,\cdots,0,1}_{m}]$,

$$E_i(t) = E_{code_i}(t). \, sum(1). \tag{7}$$

The node $a$ and node $b$ have completely opposite emotion in time $t$, when (7) satisfies both $E_a(t) = m$ and $E_b(t) = 0$ . Equivalently, if $E_a(t) = m$ indicates that node $a$ is absolutely supportive of the event, then $E_b(t) = 0$ indicates that node $b$ is absolutely against the event. When $E_c(t) = \frac{m}{2}$ or $E_c(t) = \frac{m}{2} - 1$ (m is even), $c$ is neutral to the event. In the following paper, the above principle is used as a criterion to describe the model. The ETV of each node defined as $\theta \in \Theta$, $\theta \sim M(\mu, \sigma^2)$, when $t = 0$.

The information itself also has an emotion code $E_{code\,info_n}(t)$ corresponding to m-bit $q_m$ (value 0 or 1) defined in (8), while it's Emotional Tendency Value (ETV) is defined as (9), where $E_{info_n}(t)$ represents the number of 1 in $E_{code\,info_n}(t)$.

$$E_{code\,info_n}(t) = [q_1, q_2, \cdots, q_m] \qquad (8)$$

For instance,
$$E_{code\,info_n}(t) = [\underbrace{0,1,1,0,0,\cdots,0,1}_{m}].$$

$$E_{info_n}(t) = E_{code\,info_n}(t).\,sum(1) \qquad (9)$$

The influence of information will gradually decrease along with time in the process, expressed in the (10). Moreover, there are information fragments with different emotions alternately spreading.

Not only the information itself flows between nodes, but the influence of information emotion also exists on the emotion of receiving nodes in the process. The influence is quantitatively defined as the Emotional Similarity Evaluation Function (ESEF) shown in (10). The timeline in this function will use the segmented time expression in (5). In the fractional period of the $info_n$ spreading in the network, $E_{info_n}(t = t_{info_n})$ is unique value. $E_i(t)$ indicates the Emotional Tendency Value (ETV) of node $i$ when information $info_n$ propagates to $t = t_{info_n}$.

$$\Gamma\big(E_i(t), E_{info_n}(t), t = t_{info_n}\big)$$

$$= \begin{cases} de^{\frac{\left(E_i(t) - E_{info_n}(t)\right)^2}{\sigma}} \times (1 - \vartheta t) & E_i(t) < \frac{m}{2} \\ e^{\frac{m}{2} - E_i(t)} \times de^{\frac{\left(\frac{m}{2} - E_{info_n}(t)\right)^2}{\sigma}} \times (1 - \vartheta t) & E_i(t) \geq \frac{m}{2} \end{cases} \quad E_{info_n}(t) < \frac{m}{2}$$
$$\begin{cases} de^{\frac{\left(E_i(t) - E_{info_n}(t)\right)^2}{\sigma}} \times (1 - \vartheta t) & E_i(t) > \frac{m}{2} \\ e^{E_i(t) - \frac{m}{2}} \times de^{\frac{\left(\frac{m}{2} - E_{info_n}(t)\right)^2}{\sigma}} \times (1 - \vartheta t) & E_i(t) \leq \frac{m}{2} \end{cases} \quad E_{info_n}(t) \geq \frac{m}{2}$$
$$(10)$$

Also, if $\Gamma\big(E_i(t), E_{info_n}(t), t = t_{info_n}\big) < 0$, $\Gamma\big(E_i(t), E_{info_n}(t), t = t_{info_n}\big) = 0$.

The ESEF of the same direction viewpoint is larger in (10), but the peak value of the opposite view is lower even if the information ETV is close. The detail is analyzed in first part of section 3.

The opinions from the people around us often influence our judgment, sometimes those intimate friends, and sometimes the opinions of the majority of the crowd. Abstract this phenomenon into the network, the propagating nodes will also be affected by the comprehensive emotion of the neighbors and the average emotion of the network. We define them as the neighbor affective coupling effect and the global affective coupling effect respectively, which are collectively called convergence effect. The higher the edge weight between nodes and their neighbors, the closer they are, the deeper their mutual influence, and vice versa. The definition of neighbor affective coupling effect is shown in (11).

$$M(i, t) = \frac{\sum_{Ner(i)(1)}^{Ner(i),(L)} E_{ner(i)}(t) \times \omega_{i,ner(i)}}{Num_{ner(i)}}$$
$$L = card\big(Ner(i)\big) \qquad (11)$$

Where $\omega_{i,ner}$ denotes the edge weight of node $i$ and its neighbor node $ner$, $E_{ner(i)}(t)$ is the ETV of node $i$'s neighbor node $ner$ at time $t$. The numerator in right (11) represents the weighted sum of ETV of node $i$ and its neighbor node $ner$. Node $i$ has $L = card\big(Ner(i)\big)$ neighbors. $Ner(i)(1)$ is the first neighbors of node $i$ in set $Ner(i)$. Neighbor affective coupling effect $M(i, t)$ is the weighted average sum of ETVs of node $i$ and its neighbor node $ner$.

The node emotion will be affected by the global emotion, which is often called herding.

In simpler terms, when there are more and more people with the same emotion in the whole situation, the emotion of nodes may be shaken in the convergence direction. The definition of global affective coupling effect is shown in (12).

$$\Phi(t) = \frac{\sum_{i=1}^{Num_{all}} E_i(t)}{Num_{all}} \qquad (12)$$

Where $Num_{all}$ is the total number of nodes in the network, $E_i(t)$ is the emotional value of node $i$ at time $t$, and the numerator of right equation represents the total emotional value of all nodes at time $t$. The global affective coupling effect $\Phi(t)$ is the average emotion value of all nodes in the whole network at time $t$.

In order to clarify the system of dissemination, we introduce the ESEF and convergence effect (including neighbor affective coupling effect and global affective coupling effect) into dynamic spread rate, which is defined as (13).

$$\beta\big(E_i(t), E_{info_n}(t), t = t_{info_n}\big) =$$
$$\Gamma\big(E_i(t), E_{info_n}(t), t = t_{info_n}\big) + M(i, t) + \Phi(t) \qquad (13)$$

Where the three items on the right side are emotional similarity assessment function, neighbor affective coupling effect and global affective coupling effect. It means that node $i$ has different propagation probability in different rounds of different information sources.

With the spread of multiple pieces of information with different emotion in the network, not only the state of nodes is changing, but also the emotional tendency of nodes for the whole information event is changing. So as to describe the transform of emotion, the Genetic algorithm is improved to the mutation algorithm and crossover algorithm.

Here, the emotion code $E\_code_i(t)$ of node $i$ mutates with the probability of $\sigma$ in every round $t$ of whole spread. In other word, emotional fluctuations occur in the whole process. In the mutation emotion coding, the model randomly select the mutation preselected segment, and then select the mutation sample $\{q_j\}$ with $\sigma$ probability. If $q_j = 1$ in the sample, it will change to $q_j = 0$; if $q_j = 0$, it will change to $q_j = 1$. This process is called mutation algorithm:

| **Algorithm 1: MUTATION($E\_code_i(t), \sigma$).** |
|---|
| Input: $E\_code_i(t), \sigma$ |
| Output: $E\_code_i(t)$ |
| 1:     randomly generate the list of nodes $i$ |
| 2:     len = RANDOM(0, $m$) |
| 3:     mut_list = CHOICE(0,len, $\sigma$) |
| 4:     **for** $j$ in mut_list **do** |
| 5:        **if** $E\_code_i(t)[j] == 1$ **then** |
| 6:           del $E\_code_i(t)[j]$ |
| 7:           $E\_code_i(t)$.insert($j$,0) |
| 8:        **end if** |
| 9:        **else** |
| 10:       del $E\_code_i(t)[j]$ |
| 11:       $E\_code_i(t)$.insert($j$,1) |
| 12:       **end else** |
| 13:     **end for** |

Where the second row selects a random number between 0 and m, and the third row randomly selects a column number between 0 and len with the probability of $\sigma$. The 7-th line adds 0 in the $j$-th position of the list $E\_code_i(t)$.

After the information is received by a node, the node changes the state, besides, the node emotion is affected by the information emotion. By comparing the ETV of the node with the ETV of the information, the emotional sequence of the node copy the emotional sequence of the information source unilaterally. The higher the ESEF is, the longer the copy segment sequence is. When node $i$ trigger crossover algorithm, $E_{code_i}(t)$ and $E_{code_{info_n}}(t)$ of node $i$ are the ESEF $\Gamma\big(E_i(t), E_{info_n}(t), t = t_{info_n}\big)$ calculated when the number of propagation rounds $t = t_{info_n}$. Select the cross preselected segment with length $\lceil m\Gamma \rceil$ and randomly select the starting point $d$ in the coding segment. If there is $(d + m\Gamma) < m$, the segment $\{q_d \sim q_{d+mr}\}$ in $E_{code_{info_n}}(t)$ is taken to cover the segment $E_{code_{info_n}}(t)$ in $E_{code_i}(t)$. This process implementation is described as crossover algorithm, which is the core for describing emotional swap.

| **Algorithm 2: CROSSOVER($E\_code_i(t), E_{code_{info_n}}(t)$).** |
|---|
| Input: $E\_code_i(t), E_{code_{info_n}}(t)$ |
| Output: $E\_code_i(t)$ |
| 1:   len =$\lceil \Gamma\big(E_i(t), E_{info_n}(t), t = t_{info_n}\big) \times m \rceil$ |
| 2:   start = RANDOM(0, $m - len$) |
| 3:   end = start + len |
| 4:   **if** $E\_code_i(t)$[strat, end]!=$E_{code_{info_n}}(t)$[strat, end] **then** |
| 5:     copyinfo = $E_{code_{info_n}}(t)$[strata,end] |
| 6:     **del** $E\_code_i(t)$[strata,end] |
| 7:     **for** $j$=0 to length(copyinfo) |
| 8:       $E\_code_i(t)$.insert(start+$j$, copyinfo[$j$]) |
| 9:     **end for** |
| 10:  **end if** |

The first line is the result of rounding down $\Gamma\big(E_i(t), E_{info_n}(t), t = t_{info_n}\big) \times m$, the second line selects a random number between 0 and m − len, and the eighth line adds the element copyinfo[$j$] in the start+$j$ bit of the list $E\_code_i(t)$.

## B. *Emotional Propagation Dynamic Model*

A dynamic model of information emotional propagation is established, as shown in Fig.1. Consequently, from the perspective of emotional transmission, some people who have known information will forget the news, but not forget their own emotional attitude; and other people will discuss and exchange information with each other. The node has two states: the ignorant and spreader. The ignorant refers to the node that does not know the message, while the spreader is the node that knows the message and propagates the message. The spreader will broadcast the message to neighbor node, and the infected neighbor node will change into the spreader with the spread probability $\beta$. At this time, the spreader I will change into the ignorant S with the forgetting probability $\gamma$, or into the spreader $\Gamma$ with the spread probability $\beta$ again. This model is named as the SIΓS model. In the SIΓS model, emotion is crossed in according to crossover algorithm CROSSOVER(). Here, the spreader propagates information to the ignorant or the other spreader. In simpler terms, people who already know the news will communicate with anyone. When the spreader forgets the information and transfer to the ignorant, the ETV will not be forgotten or changed, but it still may occur mutation MUTATION(). The mutation algorithm MUTATION() appears in the whole process of propagation.

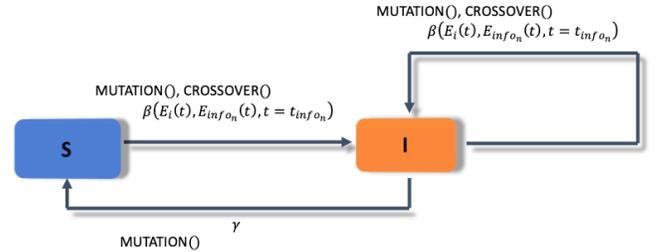

Fig. 1. Schematic diagram of hybrid spread dynamic model.

The differential equation of the SIΓS model is described as (14),

$$\begin{cases} \dfrac{dS}{dt} = \gamma I - \beta\big(E_i(t), E_{info_n}(t), t = t_{info_n}\big)SI \\ \dfrac{dI}{dt} = \beta\big(E_i(t), E_{info_n}(t), t = t_{info_n}\big)SI - \gamma I \end{cases} \quad (14)$$

By introducing (10) into (14) and (15) can be obtained,

$$\begin{cases} \dfrac{dS}{dt} = \gamma I - \left[\dfrac{\Gamma\big(E_i(t), E_{info_n}(t), t = t_{info_n} +}{M(i,t) + \Phi(t)}\right]SI \\ \dfrac{dI}{dt} = \left[\dfrac{\Gamma\big(E_i(t), E_{info_n}(t), t = t_{info_n} +}{M(i,t) + \Phi(t)}\right]SI - \gamma I \end{cases} \quad (15)$$

In (14), $S(t)$ and $I(t)$ are the proportions of the ignorant and spreader in the population respectively, where $S(t) + I(t) \equiv$

$1 (0 < s, i \le 1)$. The first term on the right side of the first equation indicates that the spreader becomes the ignorant with the probability γ. The second term on the right side of the first equation is that the ignorant becomes the spreader with the probability $\beta(E_i(t), E_{info_n}(t), t = t_{info_n})$. The initial state of the node is state S. The initial state of the node is shown in (16) and (17).

$$N = S_0 + I_0 \quad (16)$$

$$\begin{cases} S_0 = N - I_0 \\ I_0 = \beta(E_i(0), E_{info_1}(0), t = 0)N \end{cases} \quad (17)$$

$S_0$ and $I_0$ are the initial values of the ignorant and spreader respectively, where t refers to the time in (16) and (17). $N$ is the total number of nodes in the network. The right side of the second equation is the initial number of nodes when $t = 0$, that is, the dynamic propagation probability takes $\beta(E_i(0), E_{info_1}(0), t = 0)$.

## III. Mathematical Simulation

### A. Emotional Similarity Evaluation Function(ESEF)

Figures 2 represent the 3D simulation diagram of (10) $\Gamma(E_i(t), E_{info_n}(t), t = t_{info_n})$ when the parameters $E_{info_n}(t = t_{info_n})$ are 17, 7 and 21 respectively, and $T_{info_n} = [0,32]$ and $E_i(t) \in [0,32]$, where $m = 32$. Fig. 2 (a), (b) and (d) are the same graphs, which adopt three different styles to facilitate readers' better understanding. Where $P$-axis is the value of $\Gamma$, X-axis is the value of $E_i(t)$, and T is the value of $T_{info_n}$, which are the same in Fig 2 (b), (e) and (f).

In the graphs of Fig. 2 (a) and (b), the curve on $P{\sim}X(\Gamma{\sim}E_i(t))$ surface represents the ETV $\Gamma$ of $E_i(t) \in [0,32]$ when $E_{info_n}(t = t_{info_n}) = 17$. Since $E_{info_n}(t = t_{info_n}) = 17 > \frac{m}{2} = 16$, the $P{\sim}X(\Gamma{\sim}E_i(t))$ curve obeys the quasi normal distribution with $E_{info_n}(t = t_{info_n})$ as the peak value when $E_i(t) \ge 16$, and the power-law distribution with $E_{info_n}(t = t_{info_n})$ as the peak value when $E_i(t) < 16$. $\Gamma$ represents the ESEF of two emotions. The larger the value is, the more similar the two emotions are, and vice versa. The trend of $\Gamma$ is in line with the phenomenon in reality: the more extreme the emotion, the less likely to be shaken; the more neutral the idea is, the easier it is to accept various emotion, in other words, it is easier to accept various emotions. The influence of any news emotion will gradually decrease over time and eventually dissipate. The curve on $P{\sim}T(\Gamma{\sim}t_{info_n})$ surface shows that in the round number stage of information $info_n$ propagation, the ESEF of every two emotions will eventually decay to 0 with time. To put it simply, the influence of information emotion is dissipated. Therefore, the $P{\sim}T(\Gamma{\sim}t_{info_n})$ curve obeys the quasi linear distribution with the peaks of $\Gamma(E_i(0_{info_n}), E_{info_n}(0_{info_n}), 0_{info_n})$ at $T_{info_n} = [0,32]$.

In the graphs of Fig. 2 (c) and (d), since $E_{info_n}(t = t_{info_n}) = 21 > \frac{m}{2} = 16$, the principle is consistent with the three graphs in Fig. 2.

In the three graphs of Fig. 2 (e) and (f), the curve on $P{\sim}X(\Gamma{\sim}E_i(t))$ surface represents the ETV $\Gamma$ of $E_i(t) \in [0,32]$ when $E_{info_n}(t = t_{info_n}) = 7$. Since $E_{info_n}(t = t_{info_n}) = 7 < \frac{m}{2} = 16$, the $P{\sim}X(\Gamma{\sim}E_i(t))$ curve obeys the power-law distribution with $E_{info_n}(t = t_{info_n})$ as the peak value when $E_i(t) > 16$, and the quasi normal distribution with $E_{info_n}(t = t_{info_n})$ as the peak value when $E_i(t) \le 16$. Other principles are consistent with the three diagrams in Fig. 2 (a),

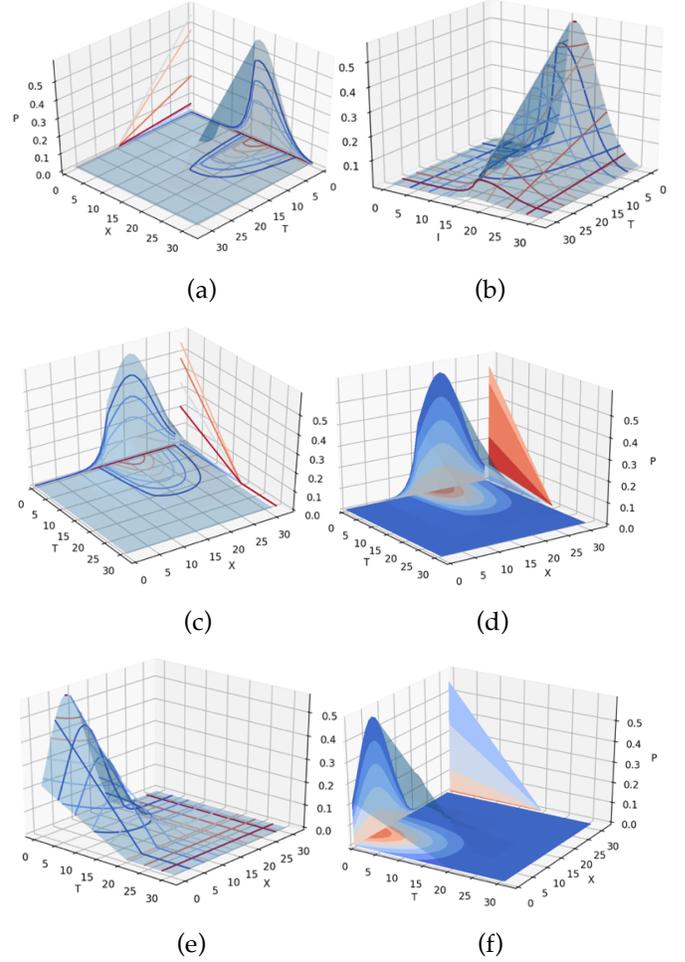

Fig. 2. ESEF (a)(b) $E_{info_n}(t_{info_n}) = 17$. (c)(d) $E_{info_n}(t_{info_n}) = 21$. (e)(f) $E_{info_n}(t_{info_n}) = 7$

(b), (c) and (d).

### B. Parameters Setup

In the mathematical simulation, the node $Num_{all} = 3000$ and the number of propagation rounds T = 180. The number of information fragments n = 7. The propagation time of each information fragment is $t_{info_1} = 30$, $t_{info_2} = 54$, $t_{info_3} = 16$, $t_{info_4} = 20$, $t_{info_5} = 19$, $t_{info_6} = 8$, $t_{info_7} = 32$. The ETV of all information segments are $E_{info_1}(t) = 21$, $E_{info_2}(t) = 17$, $E_{info_3}(t) = 1$, $E_{info_4}(t) = 6$, $E_{info_5}(t) = 19$, $E_{info_6}(t) = 9$, $E_{info_7}(t) = 17$. The value of parameter m as the length of emotion code is 32. The parameters in (10) are d = 0.67, σ = 15.7079, ϑ = 0.05. The more detail shows in Table 1.

## C. Single Information

Here we use one information fragment spread in whole process. Fig. 3 is the comprehensive analysis diagrams of emotional tendency value (ETV), when $E_{info}(t) = 11$. (a) is the stacked diagram of the accumulation of sentiment value over time. In Fig.3 (a), the horizontal axis is the propagation rounds T. When $t = 0$, the number of nodes of each ETV follows a normal distribution. (Especially, very few people have extreme emotions.) The diagram is to stack the ETV sequentially. This graph obviously indicates the number of each ETV segmentation in any propagation rounds. From the perspective of ignoring the network topology, it only counts the number of emotionally segment values in the network. (b) is a schematic diagram of node sentiment value changes over time. In Fig.3 (b), it can intuitively indicate the transform of emotion tendency value of any node in the network during the whole process of propagation. The horizontal axis is the rounds T, and the vertical axis is the number of nodes where the total number $Num_{all} = 600$. From top to bottom in the Fig. 3, as shown in the legend, they are blue $E_i(t) \in [21, 25.6]$, light blue $E_i(t) \in$ [16,21], white $E_i(t) \in [13,16]$, light red $E_i(t) \in [8,13]$, and red $E_i(t) \in [4,8]$. At the time $t = 0$ in the Fig. 3, an inflection point appears, indicating that the node has commenced to vary owing to the introduction of new information fragments. In the later stage of the spread, Fig.3 (a) evidently shows that some people with light blue emotions are transformed into light red emotions. The number of emotions with high ESEF between the information ETV gradually increases

## D. Data-driven Multi-information

So as to verify the rationality of the model, it is selected as the reference data that the trend of the poll collected by the Los Angeles times during the "2016 US election". Based on the result of mathematical simulation, the information ETV is used to model the voting mode of public opinion poll. In the way of voting, the public opinion survey includes some vote abandoning behaviors. In order to optimize visualization, we reduce the value of $E_i(t)$ by 16. In simpler terms, when $E_i(t) = 16$ ($E_i(t) = 32$ before the down regulation) indicates extreme support for Trump and extreme opposition to Clinton, and $E_i(t) = -16$ ($E_i(t) = 0$ before the down regulation) indicates extreme support for Clinton and extreme opposition to Trump. Among them, people with $E_i(t) \in [1,16]$ ($E_i(t) \in [17,32]$ before the down regulation) will definitely vote for Trump, and those with $E_i(t) = 0$ ($E_i(t) = 16$ before the down regulation) may vote for Trump or abandon their votes. People with $E_i(t) \in [-16, -2]$ ($E_i(t) \in [0,14]$ before the down regulation) will definitely vote for Clinton, and those with $E_i(t) = -1$ ($E_i(t) = 15$ before the down regulation) may vote for Clinton or abandon their votes. The number of ticket

### TABLE 1
### PARAMETERS SETUP

| Symbol | Quantity | Value |
|---|---|---|
| $T$ | Time set, (1) | - |
| $t$ | Total number of transmission rounds, (1) | 180 |
| $n$ | Number of information fragments, (3) | 7 |
| $t_{info_1}$ | Spread time of 1st information fragment, (3) | 30 |
| $t_{info_2}$ | Spread time of 2nd information fragment, (3) | 54 |
| $t_{info_3}$ | Spread time of 3rd information fragment, (3) | 16 |
| $t_{info_4}$ | Spread time of 4th information fragment, (3) | 20 |
| $t_{info_5}$ | Spread time of 5th information fragment, (3) | 19 |
| $t_{info_6}$ | Spread time of 6th information fragment, (3) | 8 |
| $t_{info_7}$ | Spread time of 7th information fragment, (3) | 32 |
| $E_{info_1}(t)$ | ETV of the 1st information fragment | 21 |
| $E_{info_2}(t)$ | ETV of the 2nd information fragment | 17 |
| $E_{info_3}(t)$ | ETV of the 3rd information fragment | 1 |
| $E_{info_4}(t)$ | ETV of the 4th information fragment | 6 |
| $E_{info_5}(t)$ | ETV of the 5th information fragment | 19 |
| $E_{info_6}(t)$ | ETV of the 6th information fragment | 9 |
| $E_{info_7}(t)$ | ETV of the 7th information fragment | 17 |
| $q_m$ | Value of the m-th bit encoding, (5) | - |
| $m$ | Emotion encoding length, (5) | 32 |
| $E_{code_i}(t)$ | Equation (7) | - |
| $E_i(t)$ | Equation (8) | - |
| $E_{code_{info_n}}(t)$ | Equation (9) | - |
| $E_{info_n}(t)$ | Equation (7) | - |
| $\Gamma$ | ESEF, (10) | - |
| $d$ | Maximum scale parameter, real reference in (10) | 0.67 |
| $\sigma$ | Real parameters in (10) | 15.71 |
| $\vartheta$ | Regression 0 scale parameter, real parameter in (10) | 0.05 |
| $M$ | Neighbor affective coupling effect (convergence effect), (11) | - |
| $Num_{ner(i)}$ | Total number of neighbor nodes of the node i, (11) | - |
| $\omega_{i,ner(i)}$ | Side rights of the node i and its neighbors, (11) | - |
| $\Phi(t)$ | Global affective coupling effect (convergence effect), (12) | - |
| $Num_{all}$ | Total number of nodes, (12) | 3000 |
| MUTATION() | mutation algorithm(math.) | - |
| CROSSOVER() | crossover algorithm(math.) | - |

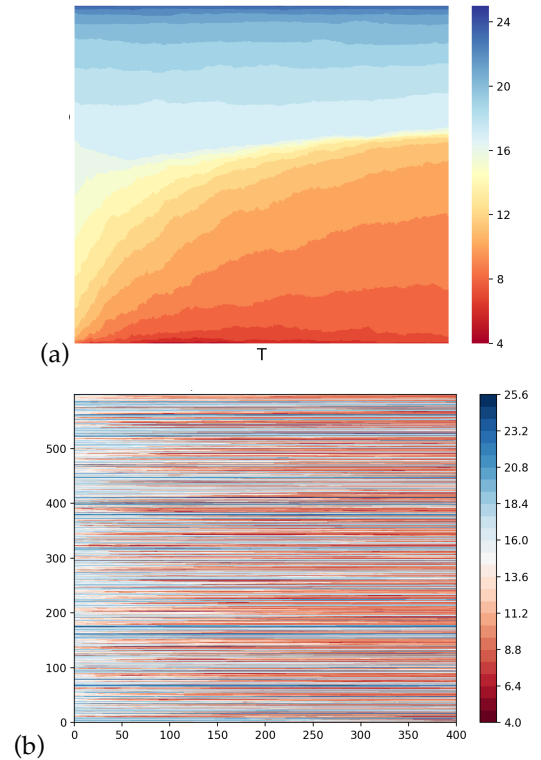

(a)

(b)

Fig. 3. The comprehensive analysis diagrams of emotional tendency value(ETV) in one information propagation. $E_{info}(t) = 11$

abandonment accounts for about 0.10% to 0.5% of the total.

Figure 8 is a voting simulation experiment based on the above settings. The two figures in Fig. 4 have orange vertical reference lines, representing events from left to right: The first dotted line: Clinton's pneumonia revealed in 2016 Sept.12; The second solid line: the first election debate in 2016; The third dotted lines: Trump's lewd remarks revealed in 2016 Oct. 7; The fourth dotted line: the second election debates in 2016 Oct.9; The fifth solid lines: the third election debate in 2016 Oct. 19; The sixth dotted line: Clinton email FBI announcement. According to the fluctuation of specific events and actual data, we quantify the emotion of information segments inserted in the process of communication as follows: the total transmission time of each information segment is $t_{info_1} = 30$, $t_{info_2} = 51$, $t_{info_3} = 84$, $t_{info_4} = 90$, $t_{info_5} = 120$, $t_{info_6} = 147$, $t_{info_7} = 158$. The ETV of all information fragments are $E_{info_1}(t) = 21$, $E_{info_2}(t) = 17$, $E_{info_3}(t) = 1$, $E_{info_4}(t) = 6$, $E_{info_5}(t) = 19$, $E_{info_6}(t) = 9$, $E_{info_7}(t) = 17$. The "Average emotion" axis of the upper subgraph in Fig. 4 represents the global emotional coupling value in the network, that is, the average value of global emotional tendency, which is represented by the black line "All nodes", where the horizontal axis T represents the number of propagation rounds. Among them, the dark thick line group is the actual data in the poll trend of the Los Angeles times during the "2016 US election". The deep red line "Trump" represents the proportion of voting for Trump, and the dark blue line "Clinton" represents that for Clinton. The light color line group is the result proportion of the simulation data voting. The light red line "Simulation Trump" represents the simulation for Trump, and the light blue line "Simulation Clinton" for Clinton.

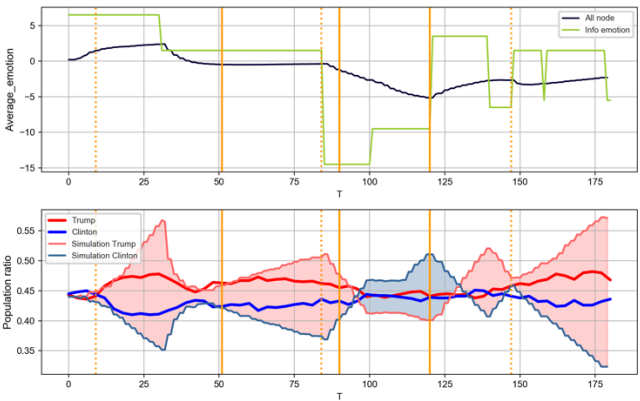

Fig. 4. Driven by data simulate information emotion dissemination.

Combined with the analysis of the upper and lower figures in Fig. 4, the influence of public opinion on voting is not highly decisive at the beginning, and there be a small shock in the figure. In election type events, people tend to have stable views, because the choice is often related to their own benefits, public opinion has tiny influence on theirs' views. Therefore, the amplitude of the actual data is slightly smaller, about 0.7% to 0.01%. It can be seen from the lower figure that the trend of the simulated data is basically consistent with the real data, and the trend is basically consistent with the trend in the upper figure.

The simulation is based on the proposed hybrid network model[38]. Fig. 5 is a rendering diagram of node emotion tendency value of a network simulation diagram. In Fig. 5, $E_i(t) = 0$ is the darkest red, and $E_i(t) = 32$ is the darkest gray. The value of $E_i(t)$ is close to the middle, the more the node dyeing tends to be white. The scale indicator bar for node coloring is shown in the Fig. 5(e). Fig. 5(a) is the network rendering diagram of the entire network node $E_i(0)$ in $t = 0$. Fig. 5(b) is the network rendering diagram of the entire network node $E_i(60)$ in $t = 60$. At this time, $E_{info_1}(0) = 21$ and $\Phi(0) = 16$ indicate that the average ETV stance of the entire network is neutral. The color in the picture is relatively neutral, and the number of gray nodes and red nodes is basically half. Fig. 5(c) is the network rendering diagram of the entire network node $E_i(120)$ in $t = 120$. At this time, $E_{info_1}(120) = 19$ and $\Phi(120) = 7$ indicate that the average ETV stance of the whole

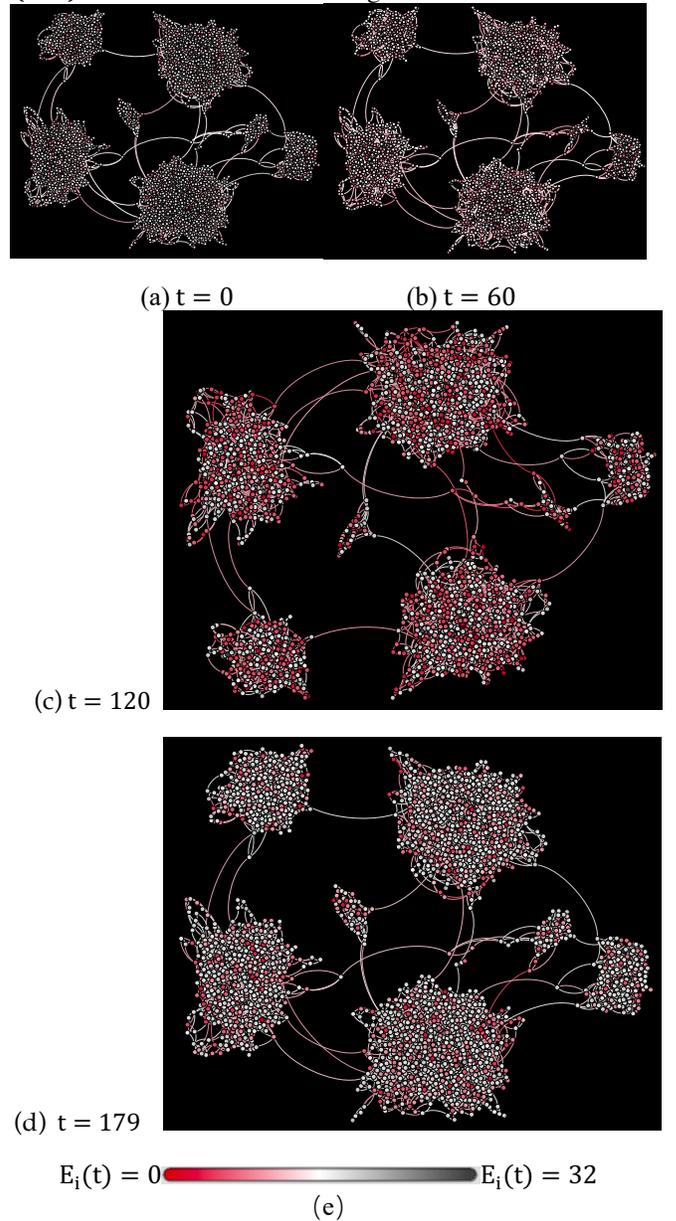

(a) t = 0      (b) t = 60

(c) t = 120

(d) t = 179

$E_i(t) = 0$    $E_i(t) = 32$
(e)

Fig. 5. Node emotion value rendering network graph.

network is biased to the position of $E_i(t) = 0$. The color in the picture is more red, the red series is much more than the gray series, and the red is darker. It also shows that this picture is the most intense emotion in the four diagrams. Fig. 5(d) is the network rendering diagram of the entire network node $E_i(179)$ in $t = 179$. At this time, $E_{info_1}(179) = 17$ and $\Phi(179) = 14$ indicate that the average ETV stance of the whole network is biased to the position of $E_i(t) = 0$. The color in the picture is redder, but the color is more neutral than Fig. 5(c), and the emotion is more neutral. Above essences also show in Fig. 4.

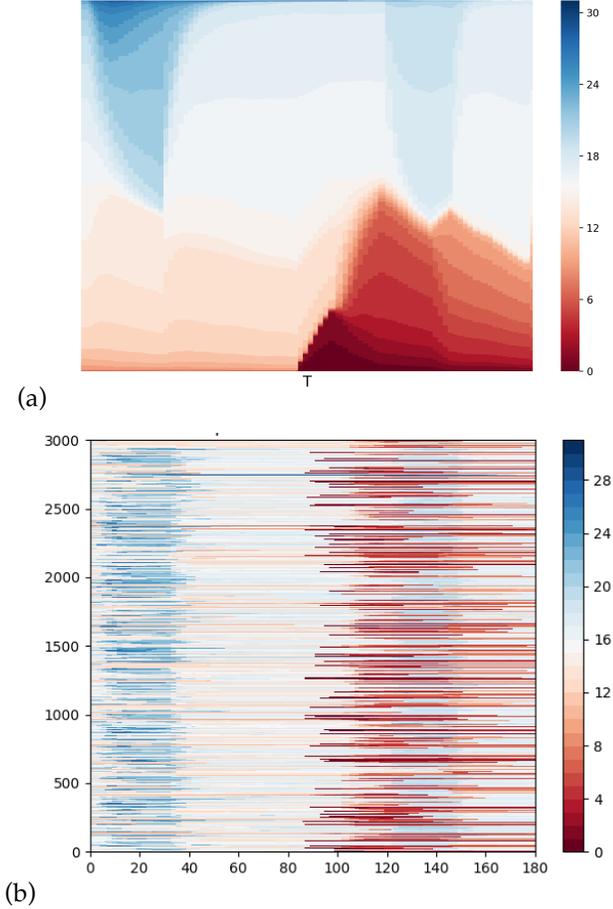

(a)

(b)

Fig. 6. A comprehensive analysis diagram of emotional tendency value(ETV) in multiple information propagation by data driven.

Fig. 6 is a comprehensive analysis diagram of emotion tendency value (ETV) in multiple emotional communication driven by real data. (a) is the stacked diagram of the accumulation of sentiment value over time. In Fig.6 (a), the horizontal axis is the propagation rounds T. When $t = 0$, the number of nodes of each ETV follows a normal distribution. (Especially, very few people have extreme emotions.) The diagram is to stack the ETV sequentially. This graph obviously indicates the number of each ETV segmentation in any propagation rounds. From the perspective of ignoring the network topology, it only counts the number of emotionally prominent segment values in the network. (b) is a schematic diagram of node sentiment value changes over time. In Fig 6 (b), it can intuitively indicate the transform of emotion tendency value of any node in the network during the whole process of propagation. The horizontal axis is the rounds T, and the vertical axis is the number of nodes where the total number $Num_{all} = 600$. From top to bottom in the Fig. 3, as shown in the legend, they are blue $E_i(t) \in [25,32]$, light blue $E_i(t) \in [18,25]$, white $E_i(t) \in [14,18]$, light red $E_i(t) \in [7,14]$, and red $E_i(t) \in [0,7]$. At the time $t = [30, 51, 84, 90, 120, 147, 158]$ in the Fig. 6, an inflection point appears, indicating that the node has commenced to vary owing to the introduction of new information fragments. Combining Fig. 4, it is clear that details of the emotional struggle between the two opposite positions.

## IV. Conclusion

In this work, the information emotion propagation model is established by introducing revamp genetic algorithms into information emotion. We simulated the interactive communication of information and emotions, and used various forms to describe the simulation process. Furthermore, genetic algorithms were used to analyze the particulars of the propagation of emotions on the basis of binary genetic coding. Finally, it presented the process model of emotion communication and game in reality. The voting mechanism of the "U.S. General Election Public Opinion Survey" was established, and simulated voting accompanied by the information and emotions propagation was conducted, and a trend similar to actual data was obtained. Our results show that the model is feasible. The controllability of emotional communication is inseparable from the dynamic interaction between groups and individuals, and it is essential for understanding the spread of public opinion and emotion.

### Acknowledgment

This research is supported by the National Natural Science Foundation of China (No. 61863025), Program for International S & T Cooperation Projects of Gansu Province (No. 144WCGA166), Program for Longyuan Young Innovation Talents and the Doctoral Foundation of LUT.

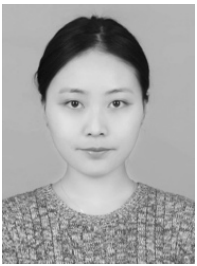

**Hongyuan Diao** received the B.S. degree in of engineering from Chongqing Normal University, Chongqing, China, in 2016, and is currently pursuing the M.S. degree in software engineering at Lanzhou University of Technology. Her main research interests include the modeling and analysis of complex networks, with applications in epidemic information and power grids.

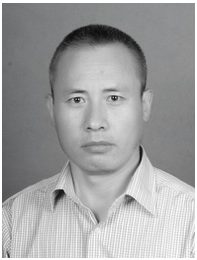

**Fuzhong Nian** received the B.S. degree of engineering from Northwest Normal University (department of Physics), LanZhou, China, in 1998, the M.S. degree in of engineering from Gansu University of Technology, Lanzhou, China, in 2004, the Ph.D. degree of engineering from Dalian University of Technology, Dalian, China, in 2011. He is interested in research at the intersection of mathematical modeling, network science, and control theory with application to biological, social and chaotic network.

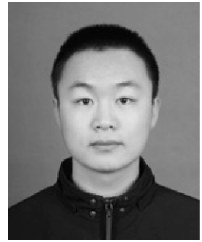

**Xuelong Yu** received the B.S. degree in of engineering from Longdong University, Qingyang, China, in 2018, and is currently pursuing the M.S. degree in software engineering at Lanzhou University of Technology. Her main research interests include the modeling and analysis of complex networks, with applications in epidemic information and power grids.

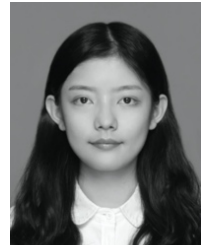

**Xirui liu** received the B.S. degree of engineering from Shaanxi Normal University, ShaanXi, China, in 2019, and is currently pursuing the M.S. degree in computer technology at Lanzhou University of Technology. Her main research interests include the modeling and analysis of complex networks, with applications in epidemic information and power grids.

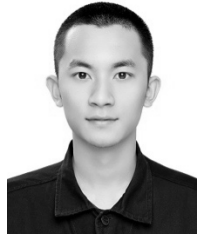

**Xinghao Liu** graduated from software engineering of Xian Xi'an University of Finance and Economics, Xi'an, China, in 2019.He is currently pursuing the M.S. degree in computer architecture with the Lanzhou University of Technology.His main research interest includes the modeling and analysis of complex networks, with applications in epidemic information the power girds.